Superconducting fluctuations and characteristic time scales in amorphous WSi

Xiaofu Zhang,[1] Adriana E. Lita,[2] Mariia Sidorova,[3] Varun B. Verma,[2] Qiang Wang,[1] Sae Woo Nam,[2] Alexej Semenov,[3] and Andreas Schilling[1]

[1]*Physics-Institute, University of Zürich, Winterthurerstrasse 190, 8057 Zürich, Switzerland*

[2]*National Institute of Standards and Technology, 325 Broadway, Boulder CO 80305, USA*

[3]*DLR Institute of Optical Systems, Rutherfordstrasse 2, 12489 Berlin, Germany*

We study magnitudes and temperature dependences of the electron-electron and electron-phonon interaction times which play the dominant role in the formation and relaxation of photon induced hotspot in two dimensional amorphous WSi films. The time constants are obtained through magnetoconductance measurements in perpendicular magnetic field in the superconducting fluctuation regime and through time-resolved photoresponse to optical pulses. The excess magnetoconductivity is interpreted in terms of the weak-localization effect and superconducting fluctuations. Aslamazov-Larkin, and Maki-Thompson superconducting fluctuation alone fail to reproduce the magnetic field dependence in the relatively high magnetic field range when the temperature is rather close to $T_c$ because the suppression of the electronic density of states due to the formation of short lifetime Cooper pairs needs to be considered. The time scale $\tau_i$ of inelastic scattering is ascribed to a combination of electron-electron ($\tau_{e-e}$) and electron-phonon ($\tau_{e-ph}$) interaction times, and a characteristic electron-fluctuation time ($\tau_{e-fl}$), which makes it possible to extract their magnitudes and temperature dependences from the measured $\tau_i$. The ratio of phonon-electron ($\tau_{ph-e}$) and electron-phonon interaction times is obtained via measurements of the optical photoresponse of WSi microbridges. Relatively large $\tau_{e-ph}/\tau_{ph-e}$ and $\tau_{e-ph}/\tau_{e-e}$ ratios ensure that in WSi the photon energy is more efficiently confined in the electron subsystem than in other materials commonly used in the technology of superconducting nanowire single-photon detectors (SNSPDs). We discuss the impact of interaction times on the hotspot dynamics and compare relevant metrics of SNSPDs from different materials.

# INTRODUCTION

In the single-photon detection process by a current-biased superconducting nanowire, the formation of the hotspot (nonequilibrium quasiparticles around the photon absorption site) and its time evolution play the most important role [1-7]. The hotspot formation can be briefly summarized as follows. (a) The incident photon is absorbed by an electron and then this highly excited electron thermalizes within a time scale of $\tau_i$ by inelastic scatterings. During this stage, a huge number of quasiparticles will be created and a hot core formed in the nanowire. (b) Nonequilibrium quasiparticles will diffuse away from the core and recombine into Cooper pairs on the characteristic time scale $\tau_0$, namely the lifetime of quasiparticles [7-9]. In other superconducting detectors, such as superconducting hot-electron bolometer [10,11], kinetic inductance detector [12,13], and superconducting tunnel junctions [14], the dynamics of the hotspot dominates detection mechanisms as well.

For the formation of the hotspot, a photon-excited electron thermalizes within a few picoseconds, depending on the details of inelastic scattering mechanisms [7,15]. It is nearly impossible to probe experimentally and distinguish these mechanisms with subpicosecond time resolution in the low temperature range. For the relaxation or cooling of the hotspot, there are different theoretical models describing this process at relatively large times [1,5,7,8,16]. In order to describe the time evolution of the hotspot completely and consistently, the perception of the characteristic time scales is necessary. In highly disordered thin superconducting films, electron-electron interaction is enhanced, and the fast inelastic scattering is mainly attributed to this interaction [15]. However, for the entire electron subsystem, energy relaxation of excited electrons occurs mainly via electron-phonon interaction [16]. Corresponding time scales, the electron-electron scattering time $\tau_{e-e}$ and the electron-phonon interaction time $\tau_{e-ph}$ play a significant role in the formation and relaxation of the hotspot.

Though the maximum count rate of a practical SNSPD is defined by its reciprocal recovering (dead) time which is controlled by the kinetic inductance of the detector [17], the time of recovery is intrinsically limited to the life-time of the hotspot [18]. As a result, the hotspot dynamics during recovering process in SNSPD sets the upper limit for the maximum count rate. It follows from simulations [19] that in conventional superconductors, e.g. Nb, the relaxation time of the hotspot is determined primarily by the temperature-dependent $\tau_{e-ph}$, i.e. hot electrons in the hotspot are cooled predominantly by the electron-phonon interaction. Although contributions of other scattering channels of electrons are less pronounced, the knowledge of temperature dependences of their characteristic time scales for different SNSPD materials is of vital importance for device design and operation. Since all these different scattering mechanisms affect the resistance in the fluctuation regime just above $T_c$, measurements of the fluctuation resistance open a channel to perceive different characteristic time scales in superconductors.

The effectiveness of photon detection by a nanowire increases with the increase in the size of the hotspot [1], and the size is larger when a larger fraction of the photon energy is confined in the electron subsystem. The relative magnitude of this fraction is called quantum yield ç. It is intuitively clear, that the quantum yield reaches maximum if the characteristic phonon-electron interaction time describing phonon re-absorption by electrons $\tau_{ph-e}$ is infinitesimal. Generally, the larger the ratio $\tau_{e-ph}/\tau_{ph-e}$, the more energy will be confined in the electron subsystem and the larger will be ç. Within the two-temperature model [20] it can be shown that for a steady-state small deviation from the equilibrium $\tau_{e-ph}/\tau_{ph-e} = C_e/C_{ph}$, and that the latter ratio can be estimated through the photoresponse of the film in the resistive state. Hence, the capacitance ratio can also be used as a criterion for device optimization. This rough consideration is consistent with the results obtained in Ref. [15] via solutions of the detailed kinetic equations for electron and phonon distribution functions.

Below we present characteristic time scales of different inelastic electron scattering processes in WSi thin films which were obtained from magnetoconductance and photoresponse measurements, and discuss their impact on the formation and relaxation of the hotspot.

**MAGNETOCONDUCTANCE**

In highly disordered films, the long inelastic lifetime of conduction electrons yields quantum interferences in a spatially extended region, which is generally called weak localization [21]. The localization effects can be directly probed by magnetotransport measurements [22]. Besides the weak localization effects, in disordered superconductors superconducting fluctuations will also significantly contribute to the total magnetoconductance. These contributions contain Aslamazov-Larkin (AL), Maki-Thompson (MT) superconducting fluctuations, fluctuations due to the suppression of the electronic density of states (DOS), and contributions from renormalization of the single-particle diffusion coefficient (DCR) [23-25]. As a result, magnetoconductance measurements in the weakly localized regime yield valuable information on intrinsic time scales of the system, e.g., the inelastic scattering time $\tau_i$, which play significant roles in the formation of the hotspot after the photon absorption. Finally, temperature dependence of $\tau_{e-ph}$ and $\tau_{e-e}$ can be obtained by analysing the different inelastic contributions to the total dephasing process.

The magnetoconductance is in most cases dominated by the weak localization effect, which is essentially caused by quantum-interference of the conduction electrons on the defects of the systems. In the two dimensional case, the conductance per sample square of weak localization effects including spin-orbit scattering and magnetic impurities scattering (neglecting the Zeeman effect in the perpendicular magnetic field) can be written as [26-28]

$$\sigma^{\mathrm{WL}}(H,T) = \sigma_0 - \frac{e^2}{2\pi^2\hbar}[\psi\left(\frac{1}{2}+\frac{1}{\omega_H\tau_1}\right) - \psi\left(\frac{1}{2}+\frac{1}{\omega_H\tau_2}\right) + \frac{1}{2}\psi\left(\frac{1}{2}+\frac{1}{\omega_H\tau_3}\right) - \frac{1}{2}\psi\left(\frac{1}{2}+\frac{1}{\omega_H\tau_4}\right)], \qquad (1)$$

where

$$\frac{1}{\tau_1} = \frac{1}{\tau_e} + \frac{1}{\tau_{so}} + \frac{1}{\tau_s},$$

$$\frac{1}{\tau_2} = \frac{4}{3}\frac{1}{\tau_{so}} + \frac{2}{3}\frac{1}{\tau_s} + \frac{1}{\tau_i} = \frac{1}{\tau_4},$$

$$\frac{1}{\tau_3} = 2\frac{1}{\tau_s} + \frac{1}{\tau_i}.$$

Here $e$ is the elementary charge, $\hbar$ is the Plank constant, $\omega_H = 4eDH/\hbar c$ is the cyclotron frequency in a disordered conductor with $D$ the diffusion constant of normal state electrons (with $D = 0.71$ and $0.85$ cm$^2$/s for 5 nm and 4 nm thick films [7]), $\tau_e$ is the elastic scattering time, $\tau_{so}$ is spin-orbit interaction time, and $\psi(x)$ is the digamma function. The parameter $\tau_s$ is the magnetic scattering time but $1/\tau_s$ is zero here because WSi is not magnetic and with no magnetic impurities. Therefore the

total excess sheet conductance due to the WL effects can be obtained by taking the zero magnetic field limit

$$\delta\sigma^{WL}(H,T) = \frac{e^2}{2\pi^2\hbar}\{\frac{3}{2}Y\left[\omega_H\left(\frac{4}{3}\frac{1}{\tau_{so}} + \frac{1}{\tau_i}\right)^{-1}\right] - \frac{1}{2}Y(\omega_H\tau_i) - Y(\omega_H\tau_1)\}, \qquad (2)$$

where $Y(x) = \psi\left(\frac{1}{2} + \frac{1}{x}\right) + \ln x$ with the limiting cases $Y(x) \approx x^2/24$ for $x \ll 1$ and for $x \gg 1$ $Y(x) \approx \ln x - 2\ln 2 - \gamma_E + \pi^2/2x$, with $\gamma_E = 0.5772$ is the Euler constant [24,29]. Moreover, since $\tau_e$ is much smaller than any other time scales here [25], the excess conductance can therefore be simplified to

$$\delta\sigma^{WL}(H,T) = \frac{e^2}{2\pi^2\hbar}\{\frac{3}{2}Y\left[\omega_H\left(\frac{4}{3}\frac{1}{\tau_{so}} + \frac{1}{\tau_i}\right)^{-1}\right] - \frac{1}{2}Y(\omega_H\tau_i)\}. \qquad (3)$$

Near the superconducting critical temperature, the total sheet resistance divergence is mainly determined by superconducting fluctuations, which cause a broad resistance transition near $T_c$. In the highly disordered superconductors, the MT fluctuation mechanism, due to coherent scattering of electrons forming Cooper pairs on impurities, describes single-particle quantum interference at impurities in the presence of superconducting fluctuations [23,30,31]. In two dimensions, the MT magnetoconductance can be written as [22]

$$\sigma^{MT} = \frac{e^2}{\pi\hbar^2}\frac{k_B T \tau_{GL}}{1 - \tau_{GL}/\tau_i}[\psi\left(\frac{1}{2} + \frac{1}{\omega_H\tau_{GL}}\right) - \psi\left(\frac{1}{2} + \frac{1}{\omega_H\tau_i}\right)]. \qquad (4)$$

Here $k_B$ is the Boltzmann constant, and $\tau_{GL}$ is the Ginzburg-Landau time ($\tau_{GL}^{-1} = \frac{8k_B T}{\pi\hbar}\ln\frac{T}{T_c}$, with $T_c = 3.9$ and $3.44$ K for 5 nm and 4 nm thick film, respectively), representing the life time of Cooper pairs, which is determined by the decay rate into two free electrons. In the zero field limit, this reduces to the well-known MT fluctuation term

$$\sigma^{MT}(H=0) = \frac{e^2}{\pi\hbar^2}\frac{k_B T \tau_{GL}}{1 - \tau_{GL}/\tau_i}\ln\frac{\tau_i}{\tau_{GL}}. \qquad (5)$$

As a result, the excess magnetoconductance due to MT fluctuation can be written as

$$\delta\sigma^{MT}(H,T) = \frac{e^2}{2\pi^2\hbar}\left[\frac{\pi^2}{4\ln\left(\frac{T}{T_c}\right)}\frac{1}{(1-\tau_{GL}/\tau_i)}\right]\cdot[Y(\omega_H\tau_{GL}) - Y(\omega_H\tau_i)]. \qquad (6)$$

The AL fluctuation contribution, which describes the effects of fluctuating Cooper pairs [22,23,32,33], is

$$\sigma^{AL} = \frac{2e^2}{\pi \hbar} \left(\frac{k_B T \tau_{GL}}{\hbar}\right) \mathcal{H}_2(\omega_H \tau_{GL}), \tag{7}$$

$$\mathcal{H}_2(x) = \frac{1}{x}\{1 - \frac{2}{x}[\psi\left(1 + \frac{1}{x}\right) - \psi\left(\frac{1}{2} + \frac{1}{x}\right)]\}. \tag{8}$$

In the zero field limit, $\mathcal{H}_2(x \to 0) \approx 1/4$, we recover from the above equation to the famous AL fluctuation conductivity [34]

$$\sigma^{AL}(H = 0) = \frac{e^2}{16\hbar} \frac{1}{\ln T/T_c}. \tag{9}$$

Finally the excess magnetoconductance can be written as

$$\delta\sigma^{AL}(H,T) = \frac{e^2}{2\pi^2 \hbar} \frac{\pi^2}{2 \ln T/T_c} [\mathcal{H}_2(\omega_H \tau_{GL}) - 0.25]. \tag{10}$$

The formation of short lifetime Cooper pairs results in a change in the number of electrons near the Fermi level. Such an indirect effect from the quasiparticles is referred to as the DOS contribution. Glatz *et al.* recently recalculated the contribution from the change of the single-particle density of states comprehensively, and in low magnetic fields near $T_c$, the DOS contribution to the conductance is [23,25]

$$\sigma^{DOS}(H,T) = \frac{14\zeta(3)e^2}{\pi^4 \hbar} [\ln\left(\omega_H \tau_{GL} \ln \frac{T}{T_c}\right) + \psi\left(\frac{1}{2} + \frac{1}{\omega_H \tau_{GL}}\right)], \tag{11}$$

where $\zeta$ is the Riemann zeta function, with $\zeta(3) = 1.202$. In the zero field limit, we have

$$\sigma^{DOS}(H = 0) = \frac{14\zeta(3)e^2}{\pi^4 \hbar} \ln(\ln T/T_c). \tag{12}$$

Therefore the excess magnetoconductance due to DOS effect can be written as

$$\delta\sigma^{DOS}(H,T) = \frac{e^2}{2\pi^2 \hbar} \frac{28\zeta(3)}{\pi^2} Y(\omega_H \tau_{GL}). \tag{13}$$

Finally, the fluctuation mechanism of renormalization of the single-particle diffusion coefficient can be neglected in the intermediate magnetic field range above $T_c$ [23,25]. In the relatively high temperature range, both AL fluctuation and the DOS contribution are dominated by the MT fluctuations [22]. However, with decreasing temperature, $\tau_{GL}$ will gradually increase and eventually exceed $\tau_i$ near $T_c$. In this case, the magnetotransport will be dominated by the AL fluctuations and DOS contribution. It should be noted here that the 2D expressions discussed above will be no longer applicable in the ultrahigh magnetic field range since the characteristic length scale $l_B = \sqrt{\hbar/2eB}$ will be lower than the film thickness $d$ [28].

Figure 1 shows the excess magnetoconductance for 5 and 4 nm thick WSi films in the relatively high temperature range, which are commonly used for SNSPD fabrications. The magnetoresistance increases with decreasing temperature and is positive in the considered magnetic field range. Above 6 K, the excess magnetoconductance can be well described by the MT fluctuation and the WL effect in the whole magnetic field range. In the low temperature range near $T_c$, the WL effect and MT fluctuation alone fail to give a satisfactory fit to the data. As a result, the excess magnetoconductance has been fitted with the WL effect and including all the superconducting fluctuation contributions, as it is shown in Fig. 2. When the temperature is relatively high, for instance as in Fig. 1, $\tau_{GL}$ is quite small and therefore $\omega_H^{-1} \gtrsim \tau_{GL}$. In these cases, the excess magnetoconductance is dominated by the MT fluctuations and can be simplified as $\delta\sigma^{MT} \propto \omega_H^2$. As a result, $\delta\sigma$ monotonically decreases with $\omega_H$, namely with the magnetic field. However, with decreasing temperature, both $\tau_{GL}$ and $\tau_i$ increase. Thus in the high magnetic field range, $\omega_H^{-1} \lesssim \tau_{GL}$, $\delta\sigma$ is found to be independent of the magnetic field. A saturation of $\delta\sigma$ will therefore appear in the high magnetic field range, as it is shown in Fig. 2. These fits yield maximum inelastic time scales $\tau_i$ of 6.6 ps for the 4 nm thick film at 4.5 K and 7.6 ps for the 5 nm thick film at 5 K.

The inelastic scattering mechanisms in the investigated temperature range mainly include electron-electron, electron-phonon, and electron-fluctuation interactions. In amorphous WSi films, the thermal diffusion length $L_T = (\hbar D/k_B T)^{1/2}$ is larger than the film thickness $d$ [35]. The electron-electron scattering rate can therefore be written as [36,37]

$$\tau_{e-e}^{-1} = \frac{e^2 R_S}{2\pi^2 \hbar} \cdot k_B T \cdot \ln\frac{\pi\hbar}{e^2 R_S}. \tag{14}$$

With respect to the electron-phonon scattering rate, we have found that $\tau_{e-ph}^{-1} \propto T^3$ [7]. Moreover, at temperatures $T$ close to $T_c$, the scattering process is dominated by superconducting fluctuations, and $\tau_{e-fl}^{-1}$ is given by [38,39]

$$\tau_{e-fl}^{-1} = \frac{e^2 R_S}{2\pi^2 \hbar} \cdot k_B T \cdot \frac{2\ln 2}{\ln\frac{T}{T_c}+D}, \quad D = \frac{4\ln 2}{\sqrt{\ln^2(\frac{\pi\hbar}{e^2 R_S})+128\hbar/e^2 R_S}-\ln^2(\frac{\pi\hbar}{e^2 R_S})}. \tag{15}$$

Figure 3 shows the best fit including the scattering mechanisms discussed above, of the total inelastic interaction time $\tau_i$. The temperature dependence of $\tau_{e-ph}$ for the 5 nm thick film is found to be $\tau_{e-ph} = \alpha \cdot T^{-3}$ with $\alpha = 5.5 \times 10^3$ ps·K$^3$, and a $\tau_{e-ph}$ = 93 ps at $T_c$ and 86 ps at 4 K. For the 4 nm thick film we find $\alpha = 4.8 \times 10^3$ ps·K$^3$, which corresponds to $\tau_{e-ph}$ = 118 ps at $T_c$ and 75 ps at 4 K. Sidorova *et al.* recently also studied the electron-phonon relaxation time in a 3.4 nm thick WSi film using an amplitude-modulated absorption of sub-THz radiation (AMAR) method, and $\tau_{e-ph}$ was estimated to be in the range of 100-200 ps at 3.4 K [40], which coincides well with our result from the magnetoresistance method. With respect to the contribution from the electron-electron interaction, a temperature dependence $\tau_{e-e} = \beta/T$ with $\beta = 95$ ps·K was determined for the 5 nm film from the fit in Fig. 3, which results in a $\tau_{e-e}$ of 24.4 ps at $T_c$. For the 4 nm thick film, we obtained $\beta = 60$ ps·K, and $\tau_{e-e}$ is found to be 17.4 ps at $T_c$.

**PHOTORESPONSE**

Microbridge from WSi film with a thickness of 5 nm was driven in the resistive state at temperatures close to $T_c$, biased with a small constant current and illuminated by subpicosecond optical pulses at the wavelength of 800 nm. The pulse energy was reduced to ensure quasi-equilibrium response that was controlled via linearity of the response magnitude versus pulse energy. The time resolution of the read-out electronics is less than 50 ps and does not affect the time evolution of the photoresponse transients at the initial stage of relaxation. In quasi-equilibrium, the photoresponse is well described by the conventional two-temperature (2-T) model [20] with the system of heat balance equations for electron and phonon subsystems,

$$\begin{cases} \frac{dT_e}{dt} = -\frac{1}{\tau_{e-ph}}(T_e - T_{ph}) + \frac{1}{C_e}P_{RF}(t) \\ \frac{dT_{ph}}{dt} = \frac{1}{\tau_{e-ph}}\frac{C_e}{C_{ph}}(T_e - T_{ph}) - \frac{1}{\tau_{esc}}(T_{ph} - T_0) \end{cases}, \quad (16)$$

where $T_e$ and $T_{ph}$ are temperatures of the electron and phonon subsystems; $T_0$ is the bath temperature; $P(t)_{RF} \propto (t/t_0)^2 e^{-mt/t_0}$ is an analytical expression describing the shape of the excitation pulse; $t_0$ ($\approx 1$ ps) is the duration of the excitation pulse; $\tau_{esc}$ is the escape time which describes cooling of the phonon subsystem via phonon escape from the film to the substrate. In the small signal regime, the photoresponse to pulsed excitation is proportional to the solution [41] of Eqs. (16) for $T_e(t)$.

Fig. 4 shows the experimental photoresponse transients for the studied microbridge and the best fit for the photoresponse at the ambient temperature of 4 K. To obtain the 2-T model fit, we solved Eq. (16) and modified the solution with the known transient function of our electric readout [41]. Because of the finite low frequency edge of the readout bandpass ($\approx 50$ MHz), the voltage transient goes below the baseline at the late stage of relaxation. This negative part of the transient is better seen on a linear scale (Fig. 4a). For the fit we used $\tau_{e-ph} = 92$ ps extracted from the magnetoconductance measurements. The fitting parameters and their best-fit values were $C_e/C_{ph} = 1.4 \pm 0.3$ and $\tau_{esc} = 190 \pm 25$ ps. The best-fit capacitance ratio agrees well with the one reported in ref. 39. A relatively large phonon escape time in ultra-thin WSi film was also reported in ref. 39 where it was associated with a significant deviation of $C_{ph}$ from the value predicted by the Debye model at low temperatures.

**DISCUSSION**

Let us now discuss parameters, which most directly affect the suitability of different superconducting materials for single-photon detection. As it was shown above, these parameters are the ratio of heat capacities of electrons and phonons, $C_e/C_{ph}$, and the ratio $\tau_{e-ph}/\tau_{e-e}$.

In WSi films, the heat capacity ratio obtained via phototresponse is by a factor of 2-3 larger than in conventional NbN films commonly used in SNSPD technology. This means that the relative amount of photon energy transferred from the absorbed photon to electrons in WSi is larger than in NbN.

Moreover, being a dirty superconductor, WSi retains the advantage of small electron diffusivity that keeps the hotspot small at the initial stage of thermalization. Furthermore, the lower rate of energy transfer from electrons to phonons $1/\tau_{e-ph}$ and the similar thermalization rate $1/\tau_{e-e}$ as compared to NbN ensure that the photon energy in WSi is for a longer time confined in the electron subsystem and allow the hotspot to grow to a larger size. Generally, materials with larger ratio $\tau_{e-ph}/\tau_{e-e}$, like WSi ($\tau_{e-ph}/\tau_{e-e}$ ~3.8 for the 5 nm films at $T_c$) [this work] or MoN ($\tau_{e-ph}/\tau_{e-e}$ ~11) [42], are more suitable for SNSPD applications when compared with conventional superconducting materials, such as NbN ($\tau_{e-ph}/\tau_{e-e}$ ~1) [42]. A further increase of this ratio can be achieved by decreasing the operation temperature, which partly explains the improved performance of SNSPD in the low temperature range. Hence, when only the efficiency and the spectral sensitivity are concerned, WSi is a better choice for SNSPD applications.

Our magnetoconductance data show that at the transition temperature the ratio $\tau_{e-ph}/\tau_{e-e}$ in the 5 nm thick WSi film is slightly larger than that in the 4 nm thick WSi film. This means that in thicker films the photon energy is more efficiently transferred to electrons. However, the larger $\tau_{e-ph}$ in the thinner films will lead to larger quasiparticles lifetimes, which makes the size of the photon-induced hotspot larger in thinner films. As a result, SNSPD based on thinner WSi films with the same wire width would extend the cut-off wavelength to longer wavelength.

The hotspot lifetime $t_{HS}$ should scale with the characteristic quasiparticle lifetime $\tau_0$, which is dependent on the critical temperature, Debye frequency and the strength of electron-phonon coupling [43]. Measurements of the lifetime of the hotspot in WSi revealed that it depends additionally on the bias current, photon energy, and the ambient temperature [6]. During the relaxation process, contributions from the bias current and Joule heat need to be considered. Moreover, the effectiveness with which phonons escape from the superconducting film should also play an important role. In relatively thick films, the relaxation rate of the phonon temperature via this channel can be described as $(T_{ph} - T_o)/\tau_{esc}^*$. Here $\tau_{esc}^* = 4\,d(A \cdot u)^{-1}$ is the bare phonon escape time which is proportional to the film thickness $d$ and is inversely proportional to the transparency $A$ of the interface between the film and the substrate for acoustic phonons and to their velocity $u$. In thin films, the relaxation of the phonon temperature slows down due to the broken isotropy of phonons and due to the restriction imposed by the film thickness on the phonon wavelengths. Though the relaxation of the phonon temperature can be still described by a single relaxation time ($\tau_{esc}$ in Eq. 16), the bare phonon escape time $\tau_{esc}^*$ does not describe the relaxation any more but is related to the phonon-electron time and the phonon bottleneck parameter $\gamma$ as $\gamma\tau_{ph-e}$. From the fitting in ref. [6], $\gamma$ is found to be around 0.3 for the thin WSi film. Using our best fit value $\tau_{ph-e} = \tau_{e-ph}(C_e/C_{ph})^{-1} = 66$ ps we estimate $\tau_{esc}^* \approx 20$ ps for the 5 nm thick film, which is consistent with the computed value 36 ps for a 3.4 nm thick film [39]. Taking all the dissipation channels into consideration, we come to the conclusion that $t_{HS}$ should not depend solely on the intrinsic quasiparticle lifetime, but is corporately determined by material parameters and the external operating conditions.

Annunziata *et al.* used the 2-T model to describe the hotspot relaxation process, and the recovery was identified by measuring the critical current $I_c(t)$ or the resistance $R(t)$ within the nanowire [19]. In the electron subsystem, relaxation is mainly determined by e-ph interaction and diffusion, while the input is provided by the Joule heat. In the phonon subsystem, phonons are mainly cooled down by the ph-e interaction, escaping to the substrate, and by diffusion. This simulation gave a good description to the latching effects in Nb and NbN SNSPDs. The authors found that the temperature dependent electron-phonon interaction time $\tau_{e-ph}$ was the dominant component in the recovery process. Hence, because of the larger $\tau_{e-ph}$, WSi based SNSPD with the same kinetic inductance as NbN based SNSPD would be more prone to latch into the resistive state after a detection event.

Though relaxation of photon-induced hotspot is affected by ambient conditions and a variety of scattering channels, in any particular material the electron-phonon interaction time defines the lifetime of quasiparticles and sets the lower limit for the lifetime of the hotspot. Generally, a faster SNSPD can be realized from the material with smaller $\tau_{e-ph}$ and larger $D$. In this case $t_{HS}$ will decrease due to the faster out-diffusion and relaxation of quasiparticles. However, a relatively shorter $\tau_{e-ph}$ value will result in a lower ς and a smaller size of the hotspot. As a result, for designing a SNSPD, a trade-off must be made between the detection efficiency and the speed of the detector.

**CONCLUSIONS**

In summary, we have found magnitudes and temperature dependences for rates of electron relaxation via different interaction channels in two-dimensional amorphous WSi films through the magnetoresistance and photoresponse measurements. The excess magnetoresistance in WSi films close to the transition temperature is well-described by AL fluctuations, MT fluctuations, and the DOS contribution. The electron-phonon interaction times provided by magnetoresistance measurements are consistent with the results obtained by absorption of amplitude-modulated sub-THz radiation and by the photoresponse to short optical pulses. In thin WSi films, an electron which has absorbed an infrared photon thermalizes via inelastic scattering within a scattering time $\tau_i \sim 7$ ps, while the electron-phonon interaction sets the lower limit for the lifetime of the hotspot to approximately 100 ps at 4 K. The relatively large $\tau_{e-ph}/\tau_{e-e} = 3.8$ and $C_e/C_{ph} = 1.4 \pm 0.3$ ratios in the 5 nm thick $W_{0.75}Si_{0.25}$ allow us to conclude that the photon energy is more efficiently transferred to electrons and confined in the electron subsystem, and that the hotspot grows to a larger size than in conventional SNSPD materials. For SNSPD applications, the material parameters of WSi result in an extended spectral range of a detector and in a larger lifetime of the radiation-induced hotspot, but increase the risk of latching.

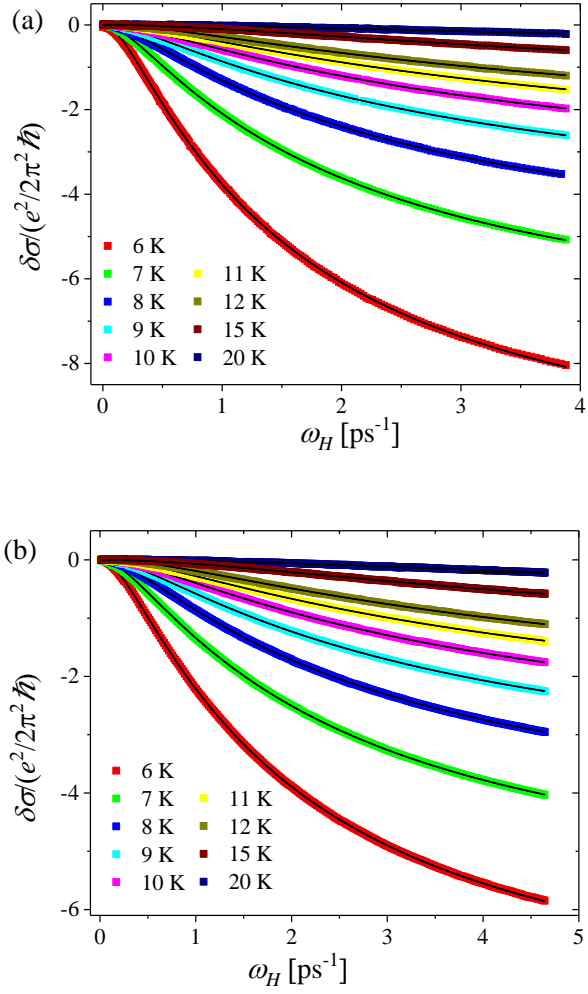

Fig. 1. The best fits of the excess magnetoconductance for 5 nm (a) and 4 nm (b) WSi films at different temperatures as specified in the legends. Fits include the WL effect and MT fluctuations as defined by Eqs. (3) and (6).

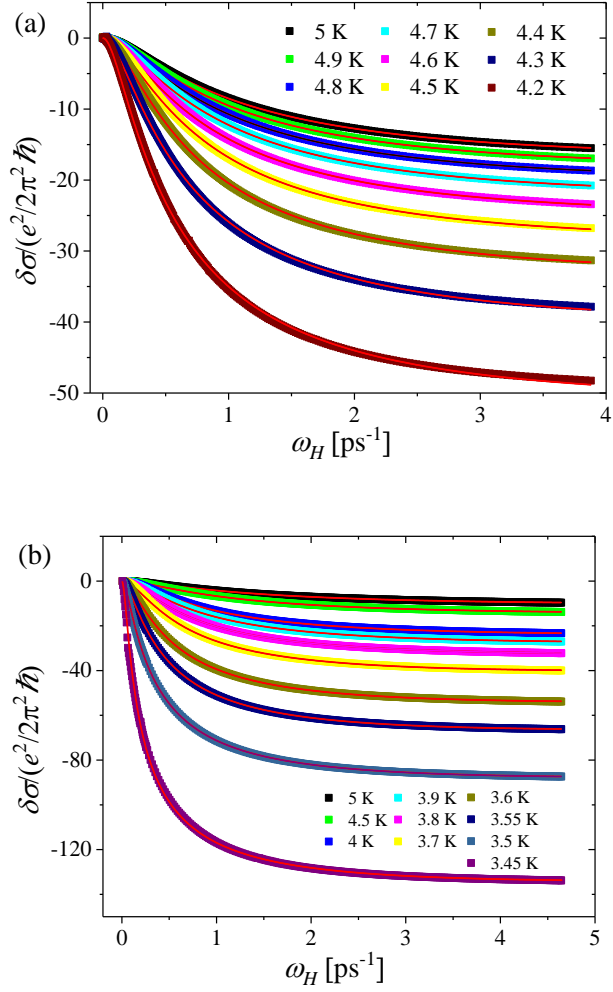

Fig. 2. The best fits of the excess magnetoconductance for 5 nm (a) and 4 nm (b) WSi films at different temperatures near $T_c$ as specified in legends. The fits consider the WL effect, MT fluctuations, AL fluctuations, and the DOS contribution as defined by Eqs. (3), (6), (10) and (13).

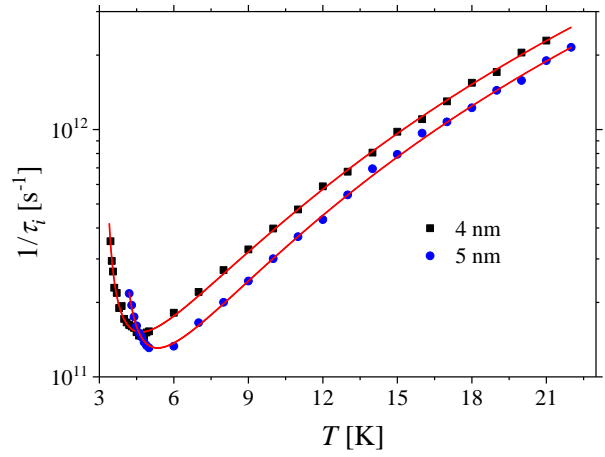

Fig. 3. Inelastic scattering rates for films with two thicknesses including e-e interaction, e-ph interaction and electron fluctuations. The solid lines correspond to best fits as explained in the text.

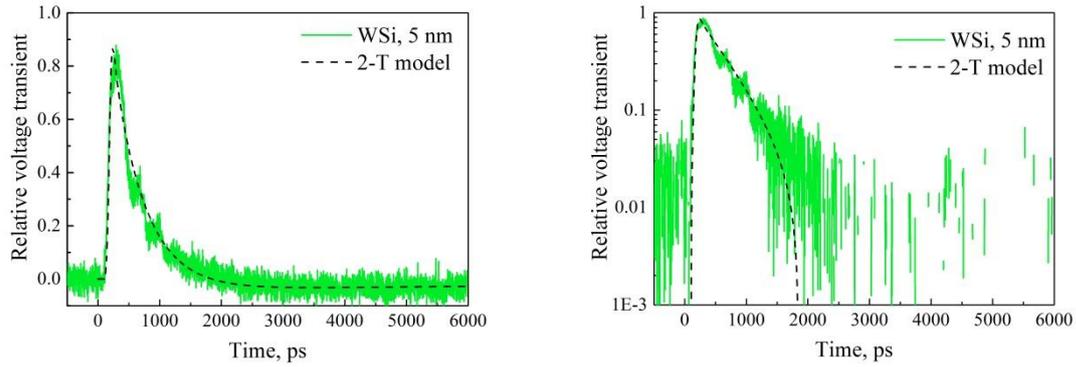

Fig. 4. Voltage photoresponse (transient) of the 5 nm thick WSi microbridgeto short optical pulse at the linear (a) and semi-logarithm (b) scales. The dashed curve represents the best fit of the response transient within the 2-T model. A few irregularities in the transient decay at times less than 1000 ps are due to signal reflections in the readout circuit.

## References

[1] A. D. Semenov, G. N. Gol'tsman, A. A. Korneev, Quantum detection by current carrying superconducting film, Phys. C (Amsterdam) **351**, 349 (2001).

[2] A. Verevkin, J. Zhang, R. Sobolewski, A. Lipatov, O. Okunev, G. Chulkova, A. Korneev, K. Smirnov, G. N. Gol'tsman, and A. Semenov, Detection efficiency of large-active-area NbN single-photon superconducting detectors in the ultraviolet to near-infrared range, Appl. Phys. Lett. **80**, 4687 (2002).

[3] A. Semenov, A. Engel, H. W. Hübers, K. Il'in, and M. Siegel, Spectral cut-off in the efficiency of the resistive state formation caused by absorption of a single-photon in current-carrying superconducting nano-strips, Eur. Phys. J. B **47**, 495 (2005).

[4] A. N. Zotova and D. Y. Vodolazov, Photon detection by current-carrying superconducting film: A time-dependent Ginzburg-Landau approach, Phys. Rev. B **85**, 024509 (2012).

[5] A. G. Kozorezov, C. Lambert, F. Marsili, M. J. Stevens, V. B. Verma, J. A. Stern, R. Horansky, S. Dyer, S. Duff, D. P. Pappas, A. Lita, M. D. Shaw, R. P. Mirin, and S. W. Nam, Quasiparticle recombination in hotspots in superconducting current-carrying nanowires, Phys. Rev. B **92**, 064504 (2015).

[6] F. Marsili, M. J. Stevens, A. Kozorezov, V. B. Verma, Colin Lambert, J. A. Stern, R. D. Horansky, S. Dyer, S. Duff, D. P. Pappas, A. E. Lita, M. D. Shaw, R. P. Mirin, and S. W. Nam, Hotspot relaxation dynamics in a current-carrying superconductor, Phys. Rev. B **93**, 094518 (2016).

[7] X. Zhang, A. Engel, Q. Wang, A. Schilling, A. Semenov, M. Sidorova, H.-W. Hübers, I. Charaev, K. Ilin, and M. Siegel, Characteristics of superconducting tungsten silicide $W_xSi_{1−x}$ for single photon detection, Phys. Rev. B **94**, 174509 (2016).

[8] A. Engel and A. Schilling, Numerical analysis of detection-mechanism models of superconducting nanowire single-photon detector, J. Appl. Phys. **114**, 214501 (2013).

[9] A. Engel, J. Lonsky, X. Zhang, and A. Schilling, Detection mechanism in SNSPD: Numerical results of a conceptually simple, yet powerful detection model, IEEE Trans. Appl. Supercond. **25**, 2200407 (2015).

[10] R. Romestain, B. Delaet, P. Renaud-Goud, I. Wang, C. Jorel, J.-C. Villegier, and J.-Ph. Poizat, Fabrication of a superconducting niobium nitride hot electron bolometer for single-photon counting, New J. Phys. **6**, 129 (2004).

[11]     D.W. Floet, E. Miedema, and T. Klapwijk, Hotspot mixing: A framework for heterodyne mixing in superconducting hot-electron bolometers, Appl. Phys. Lett. **74**, 433 (1999).

[12]     P. K. Day, H. G. LeDuc, B. A. Mazin, A. Vayonakis, and J. Zmuidzinas, A broadband superconducting detector suitable for use in large arrays, Nature (London) **425**, 817 (2003).

[13]     J. Gao, M. R. Vissers, M. O. Sandberg, F. C. S. d. Silva, S. W. Nam, D. P. Pappas, D. S. Wisbey, E. C. Langman, S. R. Meeker, B. A. Mazin, H. G. Leduc, J. Zmuidzinas, and K. D. Irwin, A titanium-nitride near-infrared kinetic inductance photon-counting detector and its anomalous electrodynamics, Appl. Phys. Lett. **101**, 142602 (2012).

[14]     A. Peacock, P. Verhoeve, N. Rando, A. van Dordrecht, B. G. Taylor, C. Erd, M. A. C. Perryman, R. Venn, J. Howlett, D. J. Goldie, J. Lumley, and M. Wallis, Single optical photon detection with a superconducting tunnel junction, Nature (London) **381**, 135 (1996).

[15]     D. Yu. Vodolazov, Single-Photon Detection by a Dirty Current-Carrying Superconducting Strip Based on the Kinetic-Equation Approach, Phys. Rev. Applied 7, 034014 (2017).

[16]     A. G. Kozorezov, A. F. Volkov, J. K. Wigmore, A. Peacock, A. Poelaert and R. den Hartog, Quasiparticle-phonon downconversion in nonequilibrium superconductors, Phys. Rev. B **61**, 11807 (2000).

[17]     A. J. Kerman, E. A. Dauler, W. E. Keicher, J. K. W. Yang, K. K. Berggren, G. Gol'tsman, and B. Voronov, Kinetic-inductance-limited reset time of superconducting nanowire photon counters, Appl. Phys. Lett. **88**, 111116 (2006).

[18]     F. Marsili, V. B. Verma, J. A. Stern, S. Harrington, A. E. Lita, T. Gerrits, I. Vayshenker, B. Baek, M. D. Shaw, R. P. Mirin, and S. W. Nam, Detecting single infrared photons with 93% system efficiency, Nat. Photonics **7**, 210 (2013).

[19]     A. J. Annunziata, O. Quaranta, D. F. Santavicca, A. Casaburi, L. Frunzio, M. Ejrnaes, M. J. Rooks, R. Cristiano, S. Pagano, A. Frydman, and D. E. Prober, Reset dynamics and latching in niobium superconducting nanowire single-photon detectors, J. Appl. Phys. **108**, 084507 (2010).

[20]     Perrin N. and Vanneste C., Response of superconducting films to a periodic optical irradiation, *Phys. Rev. B* **28**, 5150 (1983).

[21]   E. Abrahams, P. W. Anderson, D. C. Licciardello, and T. V. Ramakrishnan, Scaling theory of localization: Absence of quantum diffusion in two dimensions, Phys. Rev. Lett. **42**, 673 (1979).

[22]   A. Levchenko, Magnetoconductivity of low-dimensional disordered conductors at the onset of the superconducting transition, Phys. Rev. B **79**, 212511 (2009).

[23]   A. Glatz, A. A. Varlamov, and V. M. Vinokur, Fluctuation spectroscopy of disordered two-dimensional superconductors, Phys. Rev. B **84**, 104510 (2011).

[24]   G. M. Minkov, A. V. Germanenko, and I. V. Gornyi, Magnetoresistance and dephasing in a two-dimensional electron gas at intermediate conductances, Phys. Rev. B **70**, 245423 (2004).

[25]   A. Glatz, A. A. Varlamov, and V. M. Vinokur, Quantum fluctuations and dynamic clustering of fluctuating Cooper pairs, Europhys. Lett. **94**, 47005 (2011).

[26]   G. Bergmann, Weak localization in thin films: a time-of-flight experiment with conduction electrons, Phys. Rep. **107**, 1 (1984).

[27]   S. Maekawa and H. Fukuyama, Magnetoresistance in two-dimensional disordered systems: effects of Zeeman splitting and spin-orbit scattering, J. Phys. Soc. Jpn. **50**, 2516 (1981).

[28]   R. Rosenbaum, Superconducting fluctuations and magnetoconductance measurements of thin films in parallel magnetic fields, Phys. Rev. B **32**, 2190 (1985).

[29]   B. L. Altshuler, A. G. Aronov, A. I. Larkin, and D. E. Khmelnitskii, Anomalous magnetoresistance in semiconductors, Zh. Eksp. Teor. Fiz. **81**, 768 (1981); [Sov. Phys. JETP **54**, 411 (1981)].

[30]   R. S. Thompson, Microwave, Flux Flow, and Fluctuation Resistance of Dirty Type-II Superconductors, Phys. Rev. B **1**, 327 (1970).

[31]   J. M. B. Lopes dos Santos and E. Abrahams, Superconducting fluctuation conductivity in a magnetic field in two dimensions, Phys. Rev. B **31**, 172 (1985).

[32]   M. H. Redi, Two-dimensional fluctuation-induced conductivity above the critical temperatures, Phys. Rev. B **16**, 2027 (1977).

[33]   W. Brenig, Theory of Magnetoconductance near a Superconducting Transition in a Weakly Localized 2D Metal, J. Low Temp. Phys. **60**, 297 (1985).

[34]  L. G. Aslamazov and A. I. Larkin, Vliyanie fluktuatsii na svoistva sverkhprovodnika pri temperaturakh vyshe kriticheskoi, Fiz. Tverd. Tela (Leningrad) **10**, 1104 (1968); [Effect of Fluctuations on the Properties of a Superconductor Above the Critical Temperature, Sov. Phys. Solid State **10**, 875 (1968)].

[35]  M. Giannouri, E. Rocofyllou, C. Papastaikoudis, and W. Schilling, Weak-localization, Aslamazov-Larkin, and Maki-Thompson superconducting fluctuation effects in disordered $Zr_{1-x}Rh_x$ films above $T_c$, Phys. Rev. B **56**, 6148 (1997).

[36]  W. L. McMillan, Transition Temperature of Strong-Coupled Superconductors, Phys. Rev. **167**, 627 (1968).

[37]  B. L. Al'tshuler, A. G. Aronov, and D. E. Khmelnitsky, Effects of electron-electron collisions with small energy transfers on quantum localisation, J. Phys. C **15**, 7367 (1982).

[38]  W. Brenig, M. C. Chang, E. Abrahams, and P. Wölfle, Inelastic scattering time above the superconductivity transition in two dimensions: Dependence on disorder and magnetic field, Phys. Rev. B **31**, 7001 (1985).

[39]  W. Brenig, M. A. Paalanen, A. F. Hebard, and Wölfle, Magnetoconductance of thin-film superconductors near critical disorder, Phys. Rev. B **33**, 1691 (1986).

[40]  M. Sidorova, A. Semenov, A. Korneev, G. Chulkova, Yu. Korneeva, M. Mikhailov, A. Devizenko, A. Kozorezov, and G. Goltsman, Electron-phonon relaxation time in ultrathin tungsten silicon film, arXiv:1607.07321.

[41]  A. D. Semenov, R. S. Nebosis, Yu. P. Gousev, M. A. Heusinger, and K. F. Renk, Analysis of the nonequilibrium photoresponse of superconducting films to pulsed radiation by use of a two-temperature model, Phys. Rev. B 52(1), pp.581-590 (1995).

[42]  Y. Korneeva, I. Florya, S. Vdovichev, M. Moshkova, N. Simonov, N. Kaurova, A. Korneev, and G. Goltsman, Comparison of Hot Spot Formation in NbN and MoN Thin Superconducting Films After Photon Absorption, IEEE Trans. Appl. Supercond. **27**, 2201504 (2017).

[43] S. B. Kaplan, C. C. Chi, D. N. Langenberg, J. J. Chang, S. Jafarey, and D. J. Scalapino, Quasiparticle and phonon lifetimes in superconductors, Phys. Rev. B **14**, 4854 (1976).